\documentclass[aip,apl,reprint]{revtex4-1}

\usepackage{graphicx}
\usepackage{upgreek}
\usepackage[colorlinks=true]{hyperref}
\usepackage{tabularx}

\DeclareUnicodeCharacter{03B1}{\ensuremath{\alpha}} 

\def\U#1{{%
\def\O{\mbox{O}}
\def\u{\mbox{u}}
\mathcode`\u=\upmu
\mathcode`\O=\Omega
\mathrm{#1}}}

\def\sub#1{_{\mathrm{#1}}}

\def\AIST{Research Center for Emerging Computing Technologies, National Institute of Advanced Industrial Science and Technology~(AIST), Tsukuba, Ibaraki 305-8568, Japan}
\def\GQuAT{Global Research and Development Center for Business by Quantum-AI technology~(G-QuAT), National Institute of Advanced Industrial Science and Technology~(AIST), Tsukuba, Ibaraki 305-8568, Japan}
\def\Dev{Device Technology Research Institute, National Institute of Advanced Industrial Science and Technology~(AIST), Tsukuba, Ibaraki 305-8568, Japan}
\def\Tc{T\sub{c}}
\def\Qint{Q\sub{int}}
\def\n{\langle n\sub{ph} \rangle}

\def\aTa{$\alpha$-Ta}
\def\bTa{$\beta$-Ta}
\def\citeAY#1{\citeauthor{#1} (\citeyear{#1})}

\begin{document}


\title{Microwave characterization of tantalum superconducting resonators on silicon substrate with niobium buffer layer} 



\author{Yoshiro Urade}
\email[]{yoshiro.urade@aist.go.jp}
\affiliation{\AIST}
\affiliation{\GQuAT}

\author{Kay Yakushiji}
\affiliation{\AIST}

\author{Manabu Tsujimoto}
\affiliation{\AIST}
\affiliation{\GQuAT}

\author{Takahiro Yamada}
\affiliation{\AIST}
\affiliation{\GQuAT}

\author{Kazumasa Makise}
\altaffiliation{Present address: Advanced Technology Center, National Astronomical Observatory of Japan (NAOJ), Mitaka, Tokyo 181-8588, Japan}
\affiliation{\Dev}

\author{Wataru~Mizubayashi}
\affiliation{\AIST}
\affiliation{\GQuAT}

\author{Kunihiro Inomata}
\affiliation{\AIST}
\affiliation{\GQuAT}


\date{\today, Version 3.2}

\begin{abstract}
Tantalum thin films sputtered on unheated silicon substrates are characterized with microwaves at around $10\,\U{GHz}$ in a 10~mK environment.
We show that the phase of tantalum with a body-centered cubic lattice~(\aTa) can be grown selectively by depositing a niobium buffer layer prior to a tantalum film. The physical properties of the films, such as superconducting transition temperature and crystallinity, change markedly with the addition of the buffer layer. Coplanar waveguide resonators based on the composite film exhibit significantly enhanced internal quality factors compared with a film without the buffer layer.
The internal quality factor approaches $2\times 10^7$ at a large-photon-number limit. While the quality factor decreases at the single-photon level owing to two-level system~(TLS) loss, we have deduced that one of the causes of TLS loss is the amorphous silicon layer at the film--substrate interface, which originates from the substrate cleaning before the film deposition rather than the film itself.
The temperature dependence of the internal quality factors shows a marked rise below $200\,\U{mK}$, suggesting the presence of TLS--TLS interactions.
The present low-loss tantalum films can be deposited without substrate heating and thus have various potential applications in superconducting quantum electronics.
\end{abstract}


\maketitle 

%
%
\section{Introduction}
The integration of precisely controlled superconducting qubits has reached over 100 qubits becoming ever closer to superconducting quantum computer applications~\cite{ball_first_2021}.
This progress has been achieved through the continuous improvement in their coherence times over the past two decades in various aspects, such as circuit design~\cite{koch_charge-insensitive_2007}, geometry~\cite{barends_coherent_2013}, fabrication process~\cite{dunsworth_characterization_2017}, and materials~\cite{chang_improved_2013}.
A recent breakthrough in the material aspect is the demonstration of the long coherence times of tantalum~(Ta)-based qubits.
Transmon qubits with Ta electrodes have shown sub-millisecond energy relaxation times~\cite{place_new_2021,wang_towards_2022}.
This improvement is attributed to the dielectric loss of the Ta oxide layer formed on the surface of a Ta film being lower than that of conventionally used niobium~(Nb) films~\cite{altoe_localization_2022,verjauw_investigation_2021}.

The previous studies have revealed that it is crucial to use the phase of Ta with a body-centered cubic~(BCC) lattice (\aTa). Thin films of \aTa{} can be selectively grown on crystalline sapphire substrates at high temperatures of 500$^\circ$C~\cite{place_new_2021,li_vacuum-gap_2021} and 750$^\circ$C~\cite{crowley_disentangling_2023-1}.
However, sapphire substrates are not suitable for implementing through-substrate vias, which are crucial for three-dimensional signal routing~\cite{yost_solid-state_2020} and for suppressing parasitic electromagnetic modes in substrates~\cite{murray_predicting_2016,spring_modeling_2020,tamate_toward_2022}.
In addition, sapphire is an insulating and hard material, which complicates fabrication processes such as electron beam lithography and dicing.

Here, we focus on Ta films on silicon~(Si) substrates. Forming superconducting through-substrate vias in Si substrates is a more mature technology~\cite{yost_solid-state_2020}, and wafers with large diameters are widely available owing to the semiconductor industry. Therefore, Si is a promising base material for building large-scale superconducting quantum processors.
While \aTa{} can be grown on Si substrates by heating as with sapphire~\cite{lozano_manufacturing_2022}, high temperatures promote the formation of amorphous silicides at the Ta--Si interface, which can be an additional source of microwave loss and noise.
Moreover, a high-temperature process cannot be used after the formation of  aluminum-based Josephson junctions, which are a major component for state-of-the-art superconducting qubits.

In this study, we investigate Ta thin films sputtered on \emph{unheated} Si substrates. It is known that \aTa{} films can be grown at room temperature by using a Nb buffer layer before the deposition of Ta~\cite{face_nucleation_1987,akagi_manufacture_1993}. Nb thin films have the BCC lattice with a lattice constant close to that of \aTa{} within a $0.1\%$ mismatch, and thus they promote the nucleation and growth of \aTa{} crystals on them~\cite{face_nucleation_1987,akagi_manufacture_1993}. Microwave resonators based on such Ta films have been investigated in light of superconducting photon detectors and shown to exhibit similar loss and noise characteristics to Nb~\cite{barends_niobium_2007}, although the internal quality factor was limited up to $10^5$. Here, we revisit the room-temperature-grown Ta film and characterize it with microwave measurement setups optimized for qubit characterization. We prepare Ta films with and without a Nb buffer layer and compare their physical properties. Furthermore, we fabricate coplanar waveguide~(CPW) resonators based on the films and characterize their microwave loss in a $10\,\U{mK}$ environment from the single-photon region to the classical region. We demonstrate the potential of the present \aTa{} film as a low-loss material for use in superconducting quantum circuits.
We note that Ta-based resonators on a sapphire substrate with a Nb buffer layer have recently been investigated in Ref.~\citenum{alegria_two-level_2023}.

%
%
\section{Fabrication and Basic physical properties}
\subsection{Fabrication}
Ta and Nb films were deposited on 3-inch, double-side-polished, (001)-oriented Si wafers with a resistivity $>15\,\U{kO\cdot cm}$ and a thickness of $300\,\U{um}$ using a DC magnetron sputtering apparatus~(C-7100, Canon-Anelva). Before the deposition, the Si wafers were cleaned with 10:1 buffered hydrofluoric acid~(BHF) and argon ions to remove Si surface oxides. The thicknesses of the Ta and Nb layers were $200\,\U{nm}$ and $6\,\U{nm}$, respectively.
Then, CPW resonators were formed by photolithography and reactive ion etching with CF$_4$ gas.
Finally, the wafers were diced into $5\times 5$ $\U{mm^2}$ chips.
Each chip has four resonators capacitively coupled to a single transmission line in the hanger mode~\cite{mcrae_materials_2020}.
Their resonance frequencies are designed to be around $10$--$11\,\U{GHz}$, and the width of the center conductor and the gap between the center conductor and ground plane are $10\,\U{um}$ and $6\,\U{um}$, respectively.
For the details of the fabrication process and chip layout, see the supplementary material.

\subsection{Physical properties}\label{subsec:PP}
\begin{figure}[bt]
  \centering
  \includegraphics{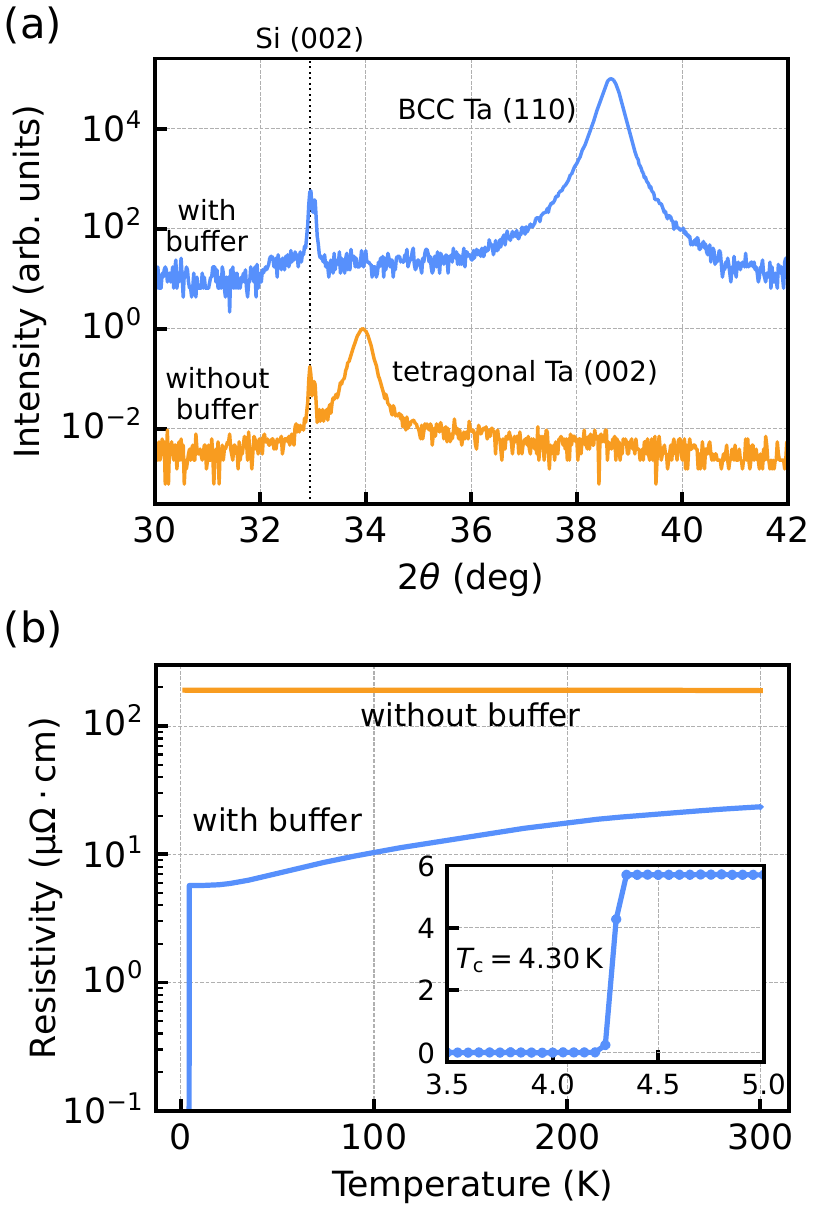}
  \caption{(a)~XRD results for the Ta films with and without the Nb buffer layer. The intensity is normalized to the maximum value. The trace of the film with the buffer layer is vertically shifted for clarity by multiplying by $10^5$. (b)~Temperature dependence of the resistivity of the Ta films with and without the Nb buffer layer from $300\,\U{K}$ to $2\,\U{K}$. The inset shows the temperature dependence of the film with the buffer layer around its superconducting transition temperature $\Tc = 4.30\,\U{K}$.\label{fig:XRD_R-T}}
 \end{figure}

We performed X-ray diffraction~(XRD) measurements of the bare films. Figure~\ref{fig:XRD_R-T}(a) shows the results.
The film without the buffer layer exhibits a peak corresponding to the tetragonal lattice of Ta (\bTa) with a rocking-curve width of $13.68^\circ$.
Meanwhile, the film with the buffer layer shows a peak corresponding to \aTa{} (110) with a rocking-curve width of $4.23^\circ$, and there is no discernible peak related to \bTa.
These results clearly show that the phase and crystallinity of the Ta layer can be controlled with the addition of the Nb buffer layer.

Next, we measured the temperature dependence of the electrical resistivity of the Ta films.
We performed four-wire measurements with Hall bar structures made of the films in a physical property measurement system~(PPMS: Model 6000, Quantum Design). Figure~\ref{fig:XRD_R-T}(b) shows the temperature dependence of the resistivity of the Ta films with and without the Nb buffer layer from $300\,\U{K}$ to $2\,\U{K}$.
The film without the buffer layer shows a high resistivity of $189\,\U{uO\cdot cm}$ and a weak temperature dependence.
The residual resistance ratio~(RRR) is $0.99$ for the resistivity at $300\,\U{K}$ and $5\,\U{K}$.
These results are consistent with the literature of \bTa{}~\cite{read_new_1965}.
In stark contrast, the film with the buffer layer shows a lower resistivity of $23.5\,\U{uO\cdot cm}$ at $300\,\U{K}$ and a higher RRR of $4.11$ and undergoes a superconducting transition at $\Tc = 4.30\,\U{K}$, which is close to the bulk $\Tc$ value of Ta~\cite{read_new_1965}.
From these observations, we conclude that high-quality \aTa{} films are grown with the buffer layer.
For simplicity, the Ta films with and without the Nb buffer layer are hereinafter referred to as \aTa{} and \bTa, respectively.

%
%

\begin{figure}[tb]
  \centering
  \includegraphics{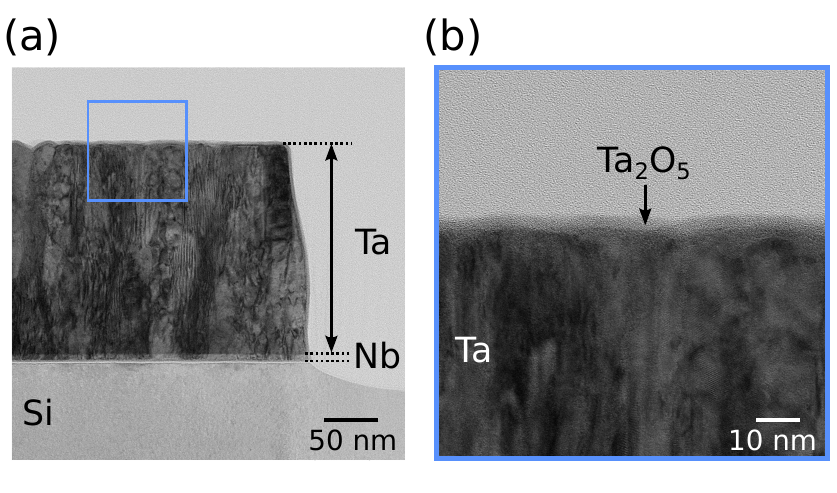}
  \caption{(a)~Cross-sectional TEM image of the center conductor of the CPW resonator made of the \aTa{} film. (b)~Magnified image of the top surface of the Ta film, indicated by the square in (a). A $2$--$3\,\U{nm}$ layer of Ta surface oxides is formed.\label{fig:TEM}}
 \end{figure}

To investigate the morphology of the \aTa{} film, we observed the cross section of the CPW resonator by transmission electron microscopy~(TEM).
Figure~\ref{fig:TEM}(a) shows a cross-sectional TEM image of the center conductor of the CPW resonator.
We can see the polycrystalline and columnar growth of Ta, as observed in the previous study~\cite{place_new_2021}.
A close-up view of the top surface of the Ta film is shown in Fig.~\ref{fig:TEM}(b).
A 2--3 nm layer of Ta surface oxides is observed, and X-ray photoelectron spectroscopy~(XPS) indicates that Ta$_2$O$_5$ is the dominant component~(see the supplementary material).

%
%
%

%
%
\section{Microwave characterization}
\subsection{Measurement setup}
The fabricated samples were placed in light-tight sample packages made of gold-plated oxygen-free copper and connected to printed circuit boards via aluminum bonding wires. The sample packages were equipped with magnetic shields and mounted on the mixing-chamber stage of a dilution refrigerator.
The complex transmission coefficient $S\sub{21}$ through the samples was repeatedly measured using a vector network analyzer while changing the probing power to sweep the average photon number in the resonators.
The temperature of the mixing-chamber stage was kept at $10$--$15\,\U{mK}$ during microwave measurements except during temperature-dependence measurements.
The internal quality factor~($Q\sub{int}$) of the resonators was extracted by fitting the theoretical curve~\cite{khalil_analysis_2012} to the measured $S\sub{21}$ spectra.
See the supplementary material for the details of the measurement setups and fittings.

%
%
\begin{figure}[h]
  \centering
  \includegraphics{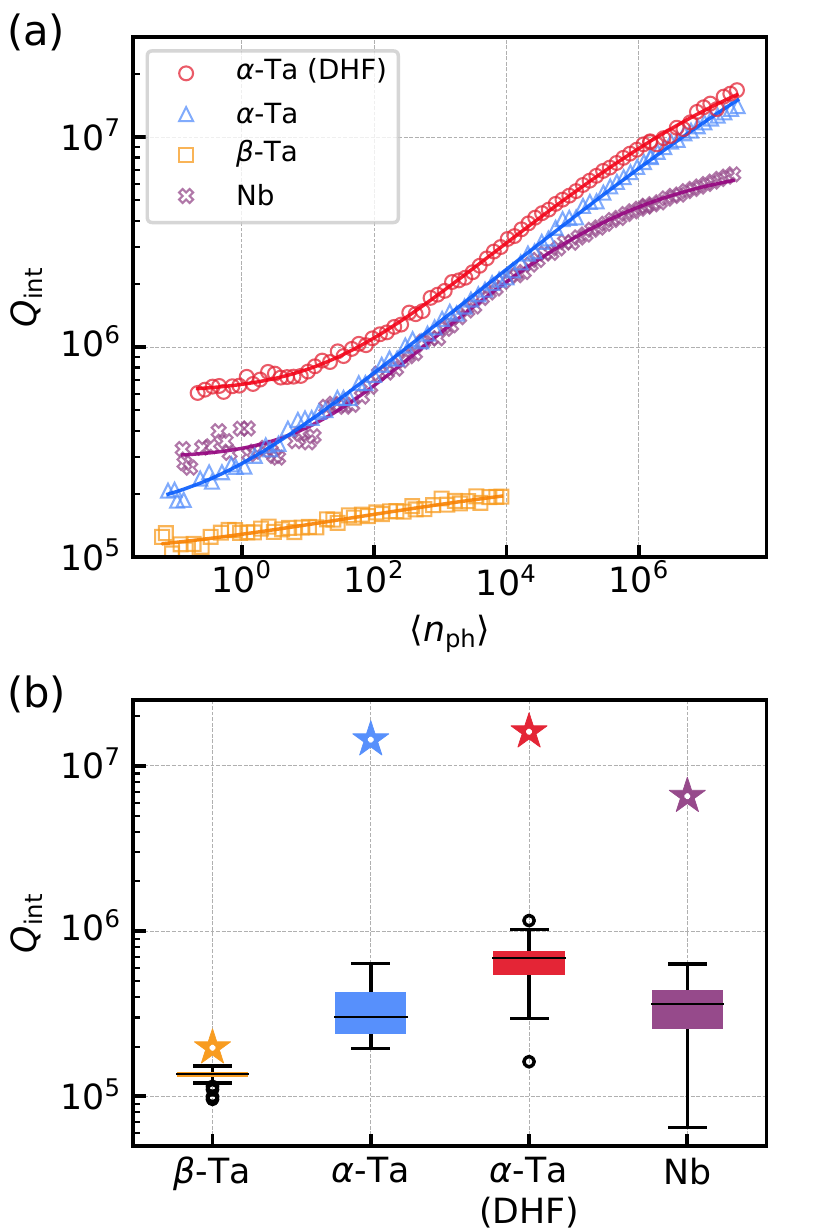}
  \caption{(a)~Internal quality factor $\Qint$ of resonator R4 versus the average photon number $\n$ in the resonator. The median values of the repeated-measurement results are plotted. The resonance frequencies of the resonators are as follows: $10.9580\,\U{GHz}$ for \aTa~(DHF), $10.9484\,\U{GHz}$ for \aTa, $11.2054\,\U{GHz}$ for Nb, and $10.8955\,\U{GHz}$ for \bTa.
  The solid lines are the fitted theoretical curves based on the TLS loss.
  (b)~Box-and-whisker plot of $\Qint$ of the resonators at $\n\approx 1$. The data sets are the fitting results of the repeated measurements of all the resonators on each chip. The upper~(lower) whisker indicates the highest~(lowest) data point in $[\mathrm{Q_1} - 1.5\times\mathrm{IQR},\mathrm{Q_3} + 1.5\times\mathrm{IQR}]$, where $\mathrm{Q_1}$ and $\mathrm{Q_3}$ are the first and third quartiles, respectively, and $\mathrm{IQR}:=\mathrm{Q_3}-\mathrm{Q_1}$. The black open circles denote the outliers. The stars indicate the median values of $\Qint$ at the maximum photon numbers~(in the case of \bTa, $\n\approx 10^4$).\label{fig:Q_R4}}
 \end{figure}

\subsection{Power dependence of $\Qint$}
Figure~\ref{fig:Q_R4}(a) shows $\Qint$ of one of the four resonators versus the average photon number $\n$ in the resonator~(see the supplementary material for the results of the other three resonators on the same chip).
Here, $\n$ is calculated from the probing power $P$ as $\n = 2Q^2P/({\hbar {\omega\sub{c}}^2 Q\sub{ext}})$~\cite{bruno_reducing_2015}, where $Q$ is the total~(loaded) quality factor; $Q\sub{ext}$ is the external~(coupling) quality factor; $\hbar$ is the reduced Planck constant; and $\omega\sub{c}$ is the angular resonance frequency.
The \aTa{} sample shows a much higher $\Qint$ than the \bTa{} sample by about two orders of magnitude at maximum, which indicates a lower residual resistance of the \aTa{} film.
Moreover, the \aTa{} sample exhibits a stronger dependence on $\n$ than does the \bTa{} sample.
This suggests that $\Qint$ of the \aTa{} sample is limited by the two-level system~(TLS) loss~\cite{martinis_decoherence_2005,gao_physics_2008, muller_towards_2019}, which saturates at large photon numbers.
In fact, the solid lines in Fig.~\ref{fig:Q_R4}(a) are fitting curves based on the following TLS loss model~\cite{wang_improving_2009,crowley_disentangling_2023-1}, demonstrating a good fit:
\begin{equation}
  \delta (\n) = \frac{F\delta_\mathrm{TLS,0}}{\sqrt{1 + \left(\frac{\n}{n\sub{sat}}\right)^{\beta_2}}} + \delta\sub{other}\label{eq:loss_power-dep},
\end{equation}
where $\delta = {\Qint}^{-1}$; $F$ is the participation ratio of the electric energy in the region where the TLS exists; $\delta_\mathrm{TLS,0}$ is the intrinsic TLS loss in the zero photon and temperature limit; $n\sub{sat}$ is the saturation photon number; $\beta_2$ is an empirical parameter that accounts for the non-uniform electric field in a CPW resonator~\cite{wang_improving_2009}; and $\delta\sub{other}$ denotes the constant loss due to other mechanisms, such as residual resistance and radiation. The free fitting parameters are $F\delta_\mathrm{TLS,0}$, $n\sub{sat}$, $\beta_2$, and $\delta\sub{other}$. See the supplementary material for the parameters of the best fit.

In marked contrast, the \bTa{} sample exhibits a weaker dependence on $\n$.
This is because another loss mechanism independent of $\n$ dominates the loss of the resonator.
The possible causes of the loss are the existence of thermal quasiparticles owing to low $\Tc$ and the residual resistance resulting from the poor crystallinity of the \bTa{} film.
Note that the data of the \bTa{} sample are not shown above $\n = 10^4$ because they exhibit distorted transmission spectra that do not fit well with the theoretical model, owing to the nonlinearity of the superconducting film~(see the supplementary material).

To investigate the origin of the TLS loss in the \aTa{} sample, we performed acid treatment~\cite{altoe_localization_2022}.
Since the surface oxides of the Si substrate are a typical source of TLS~\cite{oconnell_microwave_2008,gao_experimental_2008}, we selectively removed them using hydrofluoric acid.
We dipped an \aTa{} chip in 50:1 diluted hydrofluoric acid~(DHF) for 5 min and rinsed it with deionized water. Then, we loaded it into the dilution refrigerator and started the cooling within the same day.
The XPS measurement after the DHF treatment shows that the Si surface oxides are removed, while the Ta surface oxides still exist with slightly decreased intensity of oxidized Ta~(see the supplementary material).
As shown in Fig.~\ref{fig:Q_R4}(a), we observe an approximately twofold improvement in $\Qint$ at the single-photon level.
However, a significant $\n$ dependence remains, and thus, there are other major causes of the TLS loss in the other interfaces.

Next, to confirm whether the remaining origins of the TLS loss are related to the Ta surface oxides, we characterized Nb-based resonators.
The Nb-based samples were prepared by the same procedure as the Ta-based samples using a Nb film with a thickness of $230\,\U{nm}$.
As shown in Fig.~\ref{fig:Q_R4}(a), $\Qint$ of the Nb and \aTa{} samples exhibits a similar dependence around $\n$ of $10^0$--$10^4$.
This suggests that both samples are subject to the same causes of TLS loss.
Considering recent studies on Ta-based qubits~\cite{place_new_2021,wang_towards_2022} and Ta-encapsulated one~\cite{bal_systematic_2023}, it is expected that there appears a difference in the TLS loss of surface oxides between Ta and Nb. Therefore, it is inferred that other common causes of the TLS loss limit the quality factors of both the samples.

Another suspicious interface is the film--substrate interface.
We performed high-angle annular dark-field scanning TEM~(HAADF-STEM) and electron energy-loss spectroscopy~(EELS) at the film--substrate interface of the \aTa{} sample~(see the supplementary material). The results reveal that there are amorphous interfacial layers made up of Nb and Si with thicknesses of $1.4\,\U{nm}$ and $2.0\,\U{nm}$, respectively.
We assume that argon ion cleaning before the deposition of the films resulted in the formation of the amorphous Si layer on the substrate, and the subsequently deposited Nb layer inherited the amorphous structure in the initial stage of the film growth.
Since such an amorphous Si layer is a typical cause of TLS loss~\cite{dunsworth_characterization_2017}, we deduce that this layer partly limits $\Qint$ of the present devices. Note that there are other possible causes of the TLS loss, such as TLS in the Si bulk, surface adsorbates, and sample packaging. To decompose the loss sources, further studies with a systematic sweep of participation ratios~\cite{woods_determining_2019} are needed.

Figure~\ref{fig:Q_R4}(b) shows the $\Qint$ values of all the resonators and materials.
The box-and-whisker plot illustrates the distribution of $\Qint$ at the single-photon level~($\n\approx 1$) over the repeated measurements.
$\Qint$ of the \aTa{}-based device is limited to $0.69\times10^6$ at the single-photon level. Using the participation ratio of the film--substrate interface~$F = 0.86 \times 10^{-3}$~(calculated with the relative permittivity of Si $\varepsilon\sub{r}=11.5$)~\cite{murray_analytical_2018} and neglecting other loss causes, we can estimate the upper bound of the intrinsic loss tangent of the amorphous Si to be $1.7\times10^{-3}$ at the single-photon level, which is consistent with the results of the previous study~\cite{cicak_low-loss_2010}.
The stars in Fig.~\ref{fig:Q_R4}(b) indicate the median values of $\Qint$ at the maximum photon numbers.
Note that $\Qint$ of the \aTa{} sample is 2.5 times larger than that of the Nb sample at the large photon number.
This indicates that the non-TLS loss of Ta is smaller than that of Nb.

\subsection{Temperature dependence of $\Qint$}
%
%
\begin{figure}[tb]
  \centering
  \includegraphics{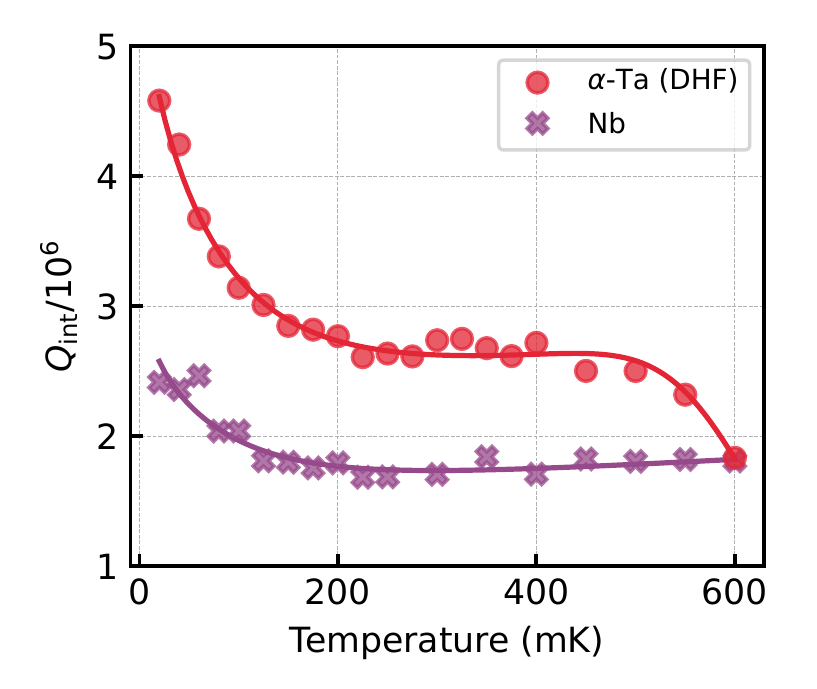}
  \caption{Temperature dependence of $\Qint$ of \aTa-based and Nb-based resonators~(R4). $\Qint$ was measured at $\n \approx 2\times 10^4$. The solid lines are the fitted theoretical curves based on the loss model incorporating the TLS--TLS interaction.\label{fig:Q-T_R4}}
 \end{figure}

We also characterized the temperature dependence of $\Qint$ to obtain further information of the TLS sources.
We controlled the temperature of the mixing chamber stage of the dilution refrigerator and measured $\Qint$ at $\n \approx 2\times 10^4$. 
Figure~\ref{fig:Q-T_R4} shows the results of the Nb sample and the \aTa{} sample with the DHF treatment.
Note that there is a rise in $\Qint$ below $200\,\U{mK}$, which cannot be explained by the widely used equation of TLS loss~\cite{mcrae_materials_2020}. A similar anomalous behavior of $\Qint$ has been discussed in Ref.~\citenum{megrant_simulating_2016}, and recently a similar temperature dependence has been clearly observed and modeled in Ref.~\citenum{crowley_disentangling_2023-1} for high-temperature-grown Ta resonators on a sapphire substrate.
The \aTa{} sample also shows a decrease in $\Qint$ above $T=400\,\U{mK}$, which is mainly attributed to thermal quasiparticles with $\Tc = 4.30\,\U{K}$ of the \aTa{} film. Note that the Nb buffer layer has a thickness of $6\,\mathrm{nm}$. Such extremely thin films generally have suppressed $\Tc$~(e.g., $\Tc \approx 6\,\U{K}$ in Refs.~\citenum{ilin_influence_2010,samsonova_signatures_2021}), and thus, the buffer layer may also contribute to the thermal quasiparticle loss.
The thick Nb resonator does not show such a behavior because of its higher $\Tc$ of $9.30\,\U{K}$.

The solid lines in Fig.~\ref{fig:Q-T_R4} are fitting curves based on the following loss model:
\begin{equation}
  \Qint = \delta^{-1} = \left(\delta\sub{TLS} + \delta\sub{QP} + \delta\sub{other} \right)^{-1},
\end{equation}
where $\delta\sub{TLS}$ and $\delta\sub{QP}$ denote the TLS loss~\cite{gao_physics_2008} and thermal quasiparticle loss~\cite{zmuidzinas_superconducting_2012}, respectively.
The TLS loss model is defined as
\begin{equation}
  \delta_\mathrm{TLS}(T) = \frac{F\delta_\mathrm{TLS,0} \tanh\left(\frac{\hbar \omega}{2 k_\mathrm{B}T}\right)}{\sqrt{1 + \left(A+{C}{T^{\beta_1}}\right)^{-1} \tanh\left(\frac{\hbar \omega}{2 k_\mathrm{B}T}\right)}}, \label{eq:delta_TLS}
\end{equation}
where $\omega$ is the angular frequency; $k_\mathrm{B}$ is the Boltzmann constant; $A$ is a constant term; $C$ is a coefficient in units of $\U{K}^{-\beta_1}$; and $\beta_1$ is an exponent accounting for the temperature dependence of the dephasing rate of the TLS ensemble due to the TLS--TLS interaction~\cite{burnett_evidence_2014,crowley_disentangling_2023-1}.
Note that we added the constant term $A$ to the model proposed in Ref.~\citenum{crowley_disentangling_2023-1} to incorporate the effect of the incomplete thermalization of the TLS ensemble.
See the supplementary material for details on the loss model and the parameters of the best fit.
The free fitting parameters are $A$, $C$, $\beta_1$, and kinetic-inductance fraction $\alpha\sub{k}$, which appears in $\delta_\mathrm{QP}$.
We determined $F\delta_\mathrm{TLS,0}$ and $\delta_\mathrm{other}$ based on the fitting results of Eq.~(\ref{eq:loss_power-dep}) with $\n$ dependence shown in Fig.~\ref{fig:Q_R4}(a).
The measured resistivity and $T\sub{c}$ in Subsec.~\ref{subsec:PP} are used in $\delta_\mathrm{QP}$.
We achieve a good fit between the theoretical curve and experimental data, and thus, it is suggested that the rise of $\Qint$ below $200\,\U{mK}$ originates from the TLS--TLS interaction.
The averaged $\bar{\beta}_1 = 1.00\pm 0.10$ and $0.81\pm0.11$ for \aTa{} and Nb, respectively, are consistent with the linear temperature dependence of the dephasing rate of the TLS ensemble predicted from the basic TLS theory~\cite{phillips_two-level_1987}.

%
%
\section{Discussion}
We compare our devices with the existing literature on Ta-based microwave resonators.
Table~\ref{tab:benchmark} presents the deposition conditions and properties of the resonators in the previous studies.
We note that although accurate comparison of the resonators is difficult due to differences in resonator geometry, measurement setups, sample packages, and so on, the comparison is useful for obtaining an overview of the state of the art.

Among room-temperature-grown devices, the resonators in this study are outstanding: $\Qint$ is one to two orders of magnitude higher.
They also exhibit a comparable performance to the Ta films grown at high temperatures, despite their relatively higher resonance frequencies, where losses tend to increase.
We think that there is still room for improvement by realizing a cleaner film--substrate interface.

The state-of-the-art results have been obtained with higher temperatures\cite{lozano_manufacturing_2022,crowley_disentangling_2023-1}~(\citeauthor{lozano_manufacturing_2022,crowley_disentangling_2023-1}), and they clearly outperform the resonators in this study. Since this can be related to not only the film but also sample packaging and resonator geometry, it is necessary to investigate them under the same condition for accurate comparison.
In addition, it is interesting to compare the uniformity of Ta phases across a wafer, since the presence of a minority fraction of \bTa{} has been reported in Ref.~\citenum{crowley_disentangling_2023-1}.

%
%
\begin{table*}[bt]
  \centering
  \caption{Comparison of Ta-based microwave resonators in the literature. $T\sub{subst}$ is the substrate temperature at deposition; $f\sub{c}$ denotes the resonance frequencies; and $Q_\mathrm{LP}$ and $Q_\mathrm{HP}$ are $\Qint$ at the lowest and highest probing powers in each study, respectively. The values of the internal quality factors are taken from the descriptions or the graphs in the references. MBE: Molecular beam epitaxy, RT: room temperature, LE: lumped element, and NA: information not available.\label{tab:benchmark}}
  \begingroup
  \setlength{\tabcolsep}{6pt}
  \renewcommand{\arraystretch}{1.3}
  \begin{tabularx}{\linewidth}{lccccccccc}
    \hline\hline
    Reference  & Method & $T\sub{subst}$ & Substrate & Buffer & RRR & Resonator & $f\sub{c}$ (GHz) & $Q\sub{LP}/10^6$  & $Q\sub{HP}/10^6$   \\ \hline
    This work  & Sputter & RT & Si (001) & Nb & 4.11 & CPW & 10--11 & 0.16--1.2 & 13--17 \\
    \citeAY{barends_niobium_2007} & Sputter & RT & Si (001) & Nb & 2.5 & CPW & 2.77 & 0.06 & 0.10\\
    \citeAY{alegria_two-level_2023} & Sputter & RT & \begin{tabular}{c}c-plane \\[-3pt] Al$_2$O$_3$\end{tabular} & Nb & 2.6 & LE & 6--8 & 0.06--0.07 & 0.10--0.13\\
    \citeAY{grigoras_qubit-compatible_2022} & Sputter & NA & Si & TiN & NA & CPW & 4--5.5 & 1.4 & 2.9\\
    \citeAY{lozano_manufacturing_2022} & Sputter & 450$^\circ\U{C}$ & Si (001) & - & NA & CPW & 4--8 & 0.83--4.8 & 8.7--55 \\
    \citeAY{shi_tantalum_2022} & Sputter & 600$^\circ\U{C}$ & \begin{tabular}{c}c-plane\\[-3pt] Al$_2$O$_3$\end{tabular} & - & 7.4 & CPW & 8.8--10.5 & 0.10--3.8 & 0.08--6.0\\
    \citeAY{alegria_two-level_2023} & Sputter & 500$^\circ\U{C}$ & \begin{tabular}{c}c-plane\\[-3pt] Al$_2$O$_3$\end{tabular} & - & 7.8 & LE & 6--8 & 0.17--0.30 & 0.45--1.8 \\
    \citeAY{jia_investigation_2023} & MBE & 550$^\circ\U{C}$ & \begin{tabular}{c}a-plane\\[-3pt] Al$_2$O$_3$\end{tabular} & - & 9.53 & CPW & 5--7 & 0.9--1.3 & 1.2--2.4\\
    \citeAY{crowley_disentangling_2023-1} & Sputter & $750^\circ\U{C}$ & Al$_2$O$_3$ & - & NA & \begin{tabular}{c}CPW,\\[-3pt] LE\end{tabular} & 4--8 & 0.1--10 & 10--200 \\
    \citeAY{jones_grain_2023} & Sputter & 500$^\circ\U{C}$ & \begin{tabular}{c}c-plane\\[-3pt] Al$_2$O$_3$\end{tabular} & - & 2.895 & CPW & 4--8 & 0.18--0.67 & 0.35--27\\
    \hline\hline
  \end{tabularx}
  \endgroup
\end{table*}

%
%
\section{Conclusion}
In conclusion, we have demonstrated that high-quality, low-loss \aTa{} thin films can be grown on unheated Si substrates by adding a Nb buffer layer.
The film with the buffer layer exhibits a superior crystallinity and a higher $\Tc$ of $4.3\,\U{K}$ and a RRR of $4.11$ than the one without the buffer layer.
The microwave resonators based on the \aTa{} film showed internal quality factors up to $2 \times 10^7$ at around $10\,\mathrm{GHz}$.
Although the loss significantly increases at the single-photon level, it is deduced that one of the origins of the loss is the amorphous Si layer between the substrate and the Nb buffer layer, which was generated by substrate cleaning before the film deposition.
The damaged interface can be easily improved by changing the initial cleaning procedure.
The other causes of the TLS loss can be, for example, bulk TLS in the Si substrate and surface adsorbates on the devices, which should be elucidated by using devices with the improved interface and sweeping participation ratios by different geometry.

The present \aTa{} film can be deposited without substrate heating, and thus, it will find a broad range of applications in superconducting quantum electronics, such as electrodes of superconducting qubits and peripheral microwave circuits.
In particular, the film is useful as a wiring layer for post-junction processes, such as the formation of air bridges~\cite{chen_fabrication_2014} and bandage metal~\cite{dunsworth_characterization_2017}, because aggressive substrate heating affects aluminum-based Josephson junctions.

\section*{Supplementary Material}
See the supplementary material for details of the fabrication process, sample design, measurement setups, measurement results, physical analyses, and loss model.

\begin{acknowledgments}
We thank Katsumi~Oono, Katsumasa~Tashiro, Go Fujii, and Shuichi~Nagasawa for their assistance in the fabrication of the samples.
We also thank Toyofumi~Ishikawa and Tomohiro~Yamaji~(NEC) for their support in the microwave measurement setup and Takayuki~Nozaki for his comments on film growth. We thank Hiroki~Kutsuma~(Tohoku University) for discussion on characterization of superconducting resonators.
This paper was based on results obtained from a project, JPNP16007, commissioned by the New Energy and Industrial Technology Development Organization~(NEDO), Japan.
The devices were fabricated in the superconducting quantum circuit fabrication facility~(Qufab) in AIST.
Part of this work was conducted at the AIST Nano-Processing Facility.
The accessible color sequence~\cite{petroff_accessible_2021} was used for data visualization in this study.
\end{acknowledgments}

\section*{Conflict of Interest}
The authors have no conflicts to disclose.

\section*{DATA AVAILABILITY}
The data that support the findings of this study are available from the corresponding author upon reasonable request.

\bibliography{Ta_CPWR_paper.bib}

\end{document}



\title{Supplementary Material for ``Microwave characterization of tantalum superconducting resonators on silicon substrate with niobium buffer layer''} 



\author{Yoshiro Urade}
\email[]{yoshiro.urade@aist.go.jp}
\affiliation{\AIST}
\affiliation{\GQuAT}

\author{Kay Yakushiji}
\affiliation{\AIST}

\author{Manabu Tsujimoto}
\affiliation{\AIST}
\affiliation{\GQuAT}

\author{Takahiro Yamada}
\affiliation{\AIST}
\affiliation{\GQuAT}

\author{Kazumasa Makise}
\altaffiliation{Present address: Advanced Technology Center, National Astronomical Observatory of Japan (NAOJ), Mitaka, Tokyo 181-8588, Japan}
\affiliation{\Dev}

\author{Wataru Mizubayashi}
\affiliation{\AIST}
\affiliation{\GQuAT}

\author{Kunihiro Inomata}
\affiliation{\AIST}
\affiliation{\GQuAT}

\date{\today, Version 2.2}

\pacs{}

\maketitle 

\section{Fabrication process}
We used 3-inch, double-side-polished, (001)-oriented Si wafers with a resistivity $>15\,\U{kO\cdot cm}$ and a thickness of $300\,\U{um}$~(Miyoshi).
First, the wafers were cleaned with 10:1 BHF~(SA, Stella Chemifa) to remove Si natural oxides and then loaded into a DC magnetron sputtering apparatus~(C-7100, Canon-Anelva).
The surfaces of the wafers were cleaned with argon ions, and then Nb and Ta films were deposited without breaking the vacuum.
The parameters of the sputtering process are detailed in Tab.~\ref{tab:sputtering}.
The pattern of the CPW resonators was formed by UV photolithography and reactive ion etching with CF$_4$ gas~(RIE-10NR, Samco). The etching parameters were as follows: CF$_4$ $50\,\U{sccm}$, pressure $5.0\,\U{Pa}$, and RF power $100\,\U{W}$, which yielded the etching rate of approximately $70\,\mathrm{nm/min}$ for both Ta and Nb films.
The remaining photoresist was removed by O$_2$ plasma ashing and N-methylpyrrolidone~(NMP).
Finally, the wafers were coated with a photoresist and diced into $5\times 5$ $\U{mm^2}$ chips, and each chip was cleaned by dipping in acetone and isopropyl alcohol baths with sonication before cryogenic measurement.

%
%
\begin{table}[htb]
     \renewcommand{\arraystretch}{1.4}
     \caption{Parameters of the sputtering process.\label{tab:sputtering}}
     \begin{tabularx}{120mm}{LCC}
          \toprule
          Parameter                     & Ta                                    & Nb                                   \\ \hline
          Target                        & $\phi 164\,\U{mm}$, t$10\,\U{mm}$, 5N & $\phi 110\,\U{mm}$, t$4\,\U{mm}$, 4N \\
          Base pressure (Pa)            & $1.5\times10^{-7}$                    & $2.2\times10^{-7}$                   \\
          Process gas                   & Kr, $15\,\U{sccm}$                    & Ar,  $60\,\U{sccm}$                  \\
          Process pressure (Pa)         & 0.05                                  & 0.1                                  \\
          Power (W)                     & 1000                                  & 500                                  \\
          T--S distance (mm)            & 240                                   & 353                                  \\
          Stage rotation (rpm)          & 100                                   & 100                                  \\
          Depo.\ rate ($\U{nm/sec}$) & 0.153                                 & 0.035                                \\
          \botrule
     \end{tabularx}
\end{table}

\section{Microwave measurement}
\subsection{Setup}
\begin{figure}[tb]
     \centering
     \includegraphics{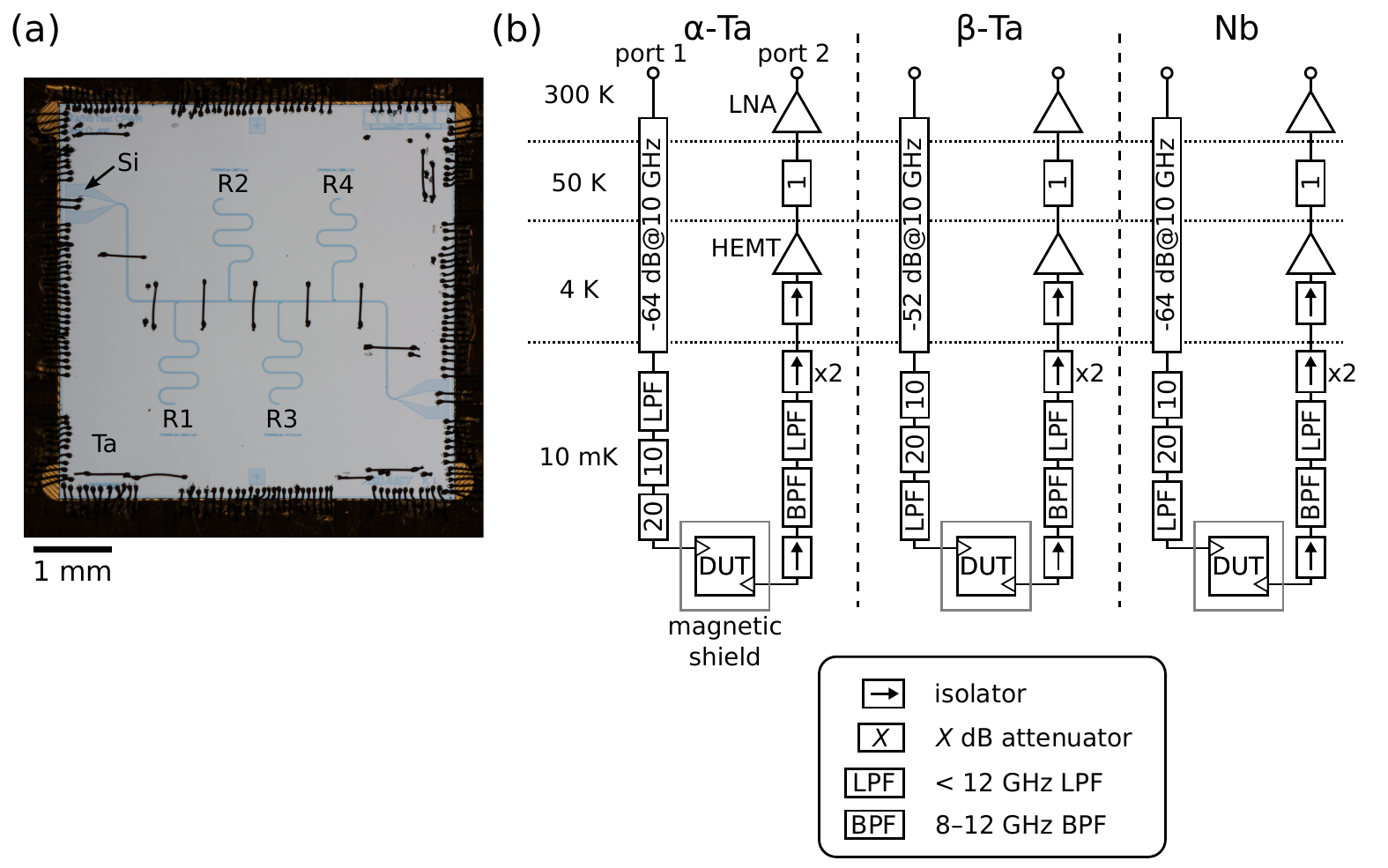}
     \caption{(a)~Micrograph of the $\alpha$-Ta resonator chip. The signal line and the ground plane are connected with the printed circuit board by aluminum wire bonding. There are four resonators, R1--R4, on the chip. (b)~Measurement setups of the samples, where LNA denotes a low-noise microwave amplifier at room temperature; HEMT denotes a cryogenic amplifier based on high-electron-mobility transistors; and LPF and BPF denote the low-pass and band-pass filters, respectively.\label{fig:setup}}
\end{figure}

A micrograph of an $\alpha$-Ta chip is shown in Fig.~\ref{fig:setup}(a).
There are four quarter-wavelength CPW resonators named R1--R4. R1 has the longest resonator length and R4 the shortest one.
To suppress the slot-line mode in the feed line, we used aluminum bonding wires to connect the separated ground planes across the center trace of the CPW.

The samples were characterized in dilution refrigerators~(LD-250, Bluefors).
The measurement setup of each sample is shown in Fig.~\ref{fig:setup}(b).
The $\beta$-Ta sample was measured in a different refrigerator system from the other samples.
At room temperature, we used vector network analyzers~(VNAs) N5232A~(for $\beta$-Ta) and M9805A~(for $\alpha$-Ta, Nb) from Keysight Technologies.
Room-temperature microwave attenuators (not shown) were also used on the input side (port 1 of the VNAs) to adjust the input microwave power.
We believe that the minor differences in the setups do not alter the conclusion of this study.

%
%
%

\subsection{Fitting}
To extract the internal quality factor $\Qint$ of the resonators, we used the following theoretical model~\cite{khalil_analysis_2012,geerlings_improving_2012} of the transmission coefficient $S\sub{21}$ for fitting with the measured data:
\begin{equation}
     S_{21}(f) = B \exp(\jj \phi\sub{off}) \left[  1 - \frac{Q/Q\sub{ext} - \mathrm{j}2Qdf/f\sub{c}}{1 + \jj2Q(f-f\sub{c})/f\sub{c}} \right], \label{eq:S21}
\end{equation}
where $f$ is the probing frequency; $B$ is the background amplitude; $\jj$ is the imaginary unit; $\phi\sub{off}$ is the phase offset; $Q$ is the total~(loaded) quality factor; $Q\sub{ext}$ is the external~(coupling) quality factor; $f\sub{c}$ is the resonance frequency; and $df$ is the parameter in units of frequency characterizing the asymmetric shape of the transmission spectrum.
From $Q$ and $Q\sub{ext}$, we obtain $\Qint$ by
\begin{equation}
     \Qint =  \left(\frac{1}{Q} - \frac{1}{Q\sub{ext}} \right)^{-1}.\label{eq:Qint}
\end{equation}

\section{Measurement data}
Here, we show the supplemental measurement data that are not shown in the main manuscript.

\subsection{Resonance frequencies of all resonators}
Table~\ref{tab:fc_all} shows the resonance frequencies of all the resonators and materials.
Because the circuit layout is identical and the depth of overetching into the Si substrate is similar, these values indicate that (i)~$\alpha$-Ta and $\beta$-Ta have similar kinetic inductances and that (ii)~Ta has a larger kinetic inductance than Nb.
It is surprising that the $\alpha$-Ta and $\beta$-Ta films exhibit similar kinetic inductances, considering the higher resistivity and lower $T\sub{c}$ of $\beta$-Ta, which generally lead to a longer penetration depth and thereby a larger kinetic inductance.

%
%
\begin{table}[htb]
     \renewcommand{\arraystretch}{1.4}
     \caption{Resonance frequencies of all the resonators and materials.\label{tab:fc_all}}
     \begin{tabularx}{110mm}{LCCCC}
          \toprule
                            & \multicolumn{4}{c}{$f\sub{c}\,(\U{GHz})$}                                                                            \\\cline{2-5}
          Material          & \multicolumn{1}{c}{R1}                    & \multicolumn{1}{c}{R2} & \multicolumn{1}{c}{R3} & \multicolumn{1}{c}{R4} \\ \hline
          $\alpha$-Ta (DHF) & 10.0028                                   & 10.3017                & 10.6197                & 10.9580                \\
          $\alpha$-Ta       & 9.9959                                    & 10.2935                & 10.6114                & 10.9484                \\
          $\beta$-Ta        & 10.0350                                   & 10.3212                & 10.5927                & 10.8955                \\
          Nb                & 10.2293                                   & 10.5350                & 10.8598                & 11.2054                \\ \botrule
     \end{tabularx}
\end{table}

\subsection{$\Qint$ versus $\n$ and $T$ of the other resonators}
\begin{figure}[h]
     \centering
     \includegraphics{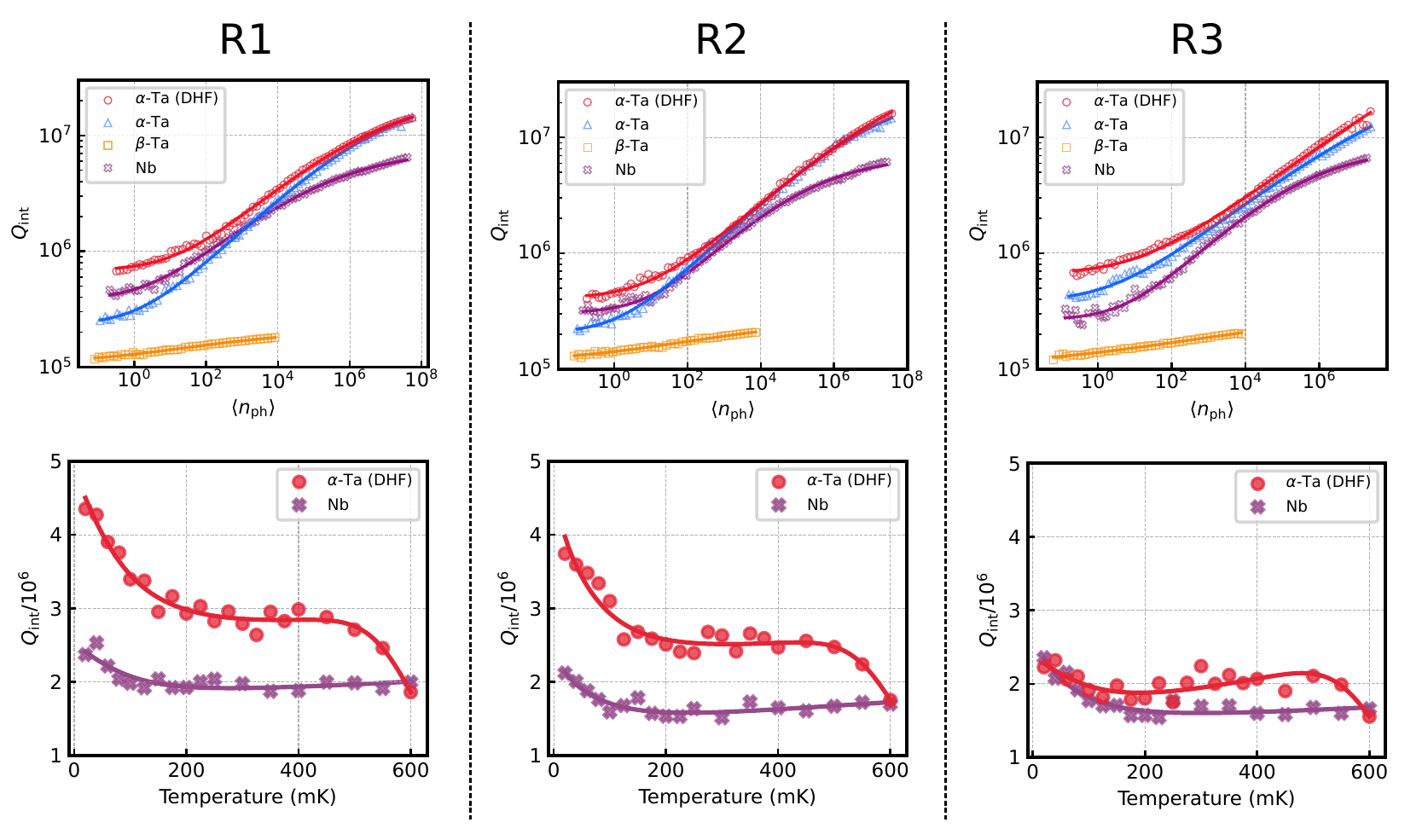}
     \caption{(top)~Internal quality factor $\Qint$ of resonators R1--R3 versus the average photon number $\n$. The median values of the repeated measurements are plotted. The solid lines are the fitted theoretical curves based on the loss model in Eq.~(1) of the main manuscript. (bottom)~Temperature dependence of $\Qint$ of the $\alpha$-Ta-based and Nb-based resonators R1--R3. $\Qint$ was measured at $\n \approx 2\times 10^4$. The solid lines are the fitted theoretical curves based on the loss model in Eq.~(\ref{eq:delta_tot}). \label{fig:Q_all}}
\end{figure}

The $\Qint$ versus $\n$ plot in Fig.~3(a) of the main manuscript exhibits the data of resonator R4.
Figure~\ref{fig:Q_all} shows the corresponding plots of resonators R1--R3.
The fitting parameters of the best-fit curves are shown in Tab.~\ref{tab:fit_Q-photon}.
The temperature dependence of $\Qint$ corresponding to Fig.~4 of the main manuscript is also shown in the bottom row of Fig.~\ref{fig:Q_all}.
The fitting parameters of the best-fit curves are summarized in Tab.~\ref{tab:fit}.

%
%
%

%
%
\begin{table}[h]
     \renewcommand{\arraystretch}{1.4}
     \caption{Fitting parameters obtained by fitting Eq.~(1) of the main manuscript with the $\n$ dependence of $\Qint$. The errors indicate one standard deviation of the parameter estimation.\label{tab:fit_Q-photon}}
     \begin{tabularx}{160mm}{RCCCC}
          \toprule
          Resonator                 & $\delta\sub{other}\,(10^{-8})$ & $F\delta_\mathrm{TLS,0}\,(10^{-6})$ & $n\sub{sat}$       & $\beta_2$ \\ \hline
          $\alpha$-Ta (DHF)\quad R1 & $4.5 \pm 0.2$  & $1.44 \pm 0.03$                    & $15 \pm 2$ & $0.539 \pm 0.009$ \\
          R2                        & $2.3 \pm 0.3$  & $2.44 \pm 0.05$                  & $8.2 \pm 1.3$ & $0.545 \pm 0.009$ \\
          R3                        & $0.5 \pm 0.6$    & $1.52 \pm 0.05$                    & $13 \pm 3$ & $0.457 \pm 0.014$ \\
          R4                        & $3.0 \pm 0.2$  & $1.61 \pm 0.02$                    & $20 \pm 2$ & $0.549 \pm 0.008$ \\
          $\alpha$-Ta\quad R1       & $4.41 \pm 0.15$    & $4.29 \pm 0.07$                    & $1.5 \pm 0.2$ & $0.587 \pm 0.005$ \\
          R2                        & $3.9 \pm 0.3$  & $4.90 \pm 0.14$                    & $1.4 \pm 0.2$ & $0.602 \pm 0.008$ \\
          R3                        & $2.7 \pm 0.5$    & $2.62 \pm 0.08$                    & $2.4 \pm 0.6$ & $0.481 \pm 0.010$ \\
          R4                        & $1.4 \pm 0.3$  & $6.2 \pm 0.2$                   & $0.25 \pm 0.06$ & $0.512 \pm 0.007$ \\
          $\beta$-Ta\quad R1        & $(4.6 \pm 0.3)\times 10^2$    & $4.6 \pm 0.7$                    & $0.7 \pm 0.6$ & $0.33 \pm 0.06$ \\
          R2                        & $(3.3 \pm 0.8)\times 10^2$  & $5.4 \pm 1.5$                    & $0.9 \pm 1.1$ & $0.27 \pm 0.09$ \\
          R3                        & $(2.6 \pm 1.2)\times 10^2$    & $7 \pm 2$                    & $0.3 \pm 0.6$ & $0.21 \pm 0.08$ \\
          R4                        & $(3.5 \pm 1.7)\times 10^2$  & $7 \pm 4$                   & $0.3 \pm 1.0$ & $0.27 \pm 0.17$ \\
          Nb\quad R1                & $13.0 \pm 0.4$    & $2.55 \pm 0.07$                    & $2.4 \pm 0.5$ & $0.525 \pm 0.010$ \\
          R2                        & $14.3 \pm 0.5$  & $3.15 \pm 0.07$                    & $8.8 \pm 1.3$ & $0.624 \pm 0.013$ \\
          R3                        & $12.2 \pm 0.6$    & $3.66 \pm 0.09$                    & $5.9 \pm 1.0$ & $0.620 \pm 0.014$ \\
          R4                        & $13.1 \pm 0.7$  & $3.23 \pm 0.10$                    & $10 \pm 2$ & $0.633 \pm 0.019$ \\
          \botrule
     \end{tabularx}
\end{table}

%
%
\begin{table}[h]
     \renewcommand{\arraystretch}{1.4}
     \caption{Fitting parameters obtained by fitting Eq.~(\ref{eq:delta_tot}) with the temperature dependence of $\Qint$. The errors indicate one standard deviation of the parameter estimation.\label{tab:fit}}
     \begin{tabularx}{150mm}{RCCCC}
          \toprule
          Resonator                 & $A\,(10^{-3})$ & $C\,(10^{-1}\,\U{K}^{-\beta_1})$ & $\beta_1$       & $\alpha\sub{k}$ \\ \hline
          $\alpha$-Ta (DHF)\quad R1 & $13 \pm 2$  & $2.5 \pm 0.3$                    & $1.16 \pm 0.11$ & $0.167 \pm 0.012$ \\
          R2                        & $6 \pm 2$  & $0.91 \pm 0.10$                  & $0.90 \pm 0.13$ & $0.16 \pm 0.02$ \\
          R3                        & $(8 \pm 2)\times 10^1$    & $4.3 \pm 1.1$                    & $0.98 \pm 0.29$ & $0.19 \pm 0.03$ \\
          R4                        & $9 \pm 2$  & $2.08 \pm 0.14$                    & $0.96 \pm 0.07$ & $0.151 \pm 0.009$ \\
          Nb\quad R1                & $11 \pm 2$     & $0.87 \pm 0.11$                    & $0.99 \pm 0.16$ & N/A             \\
          R2                        & $6 \pm 4$  & $0.70 \pm 0.06$                  & $0.69 \pm 0.15$ & N/A             \\
          R3                        & $4 \pm 2$  & $0.62 \pm 0.06$                  & $0.80 \pm 0.13$ & N/A             \\
          R4                        & $3 \pm 2$  & $0.60 \pm 0.05$                  & $0.74 \pm 0.12$ & N/A             \\
          \botrule
     \end{tabularx}
\end{table}

\subsection{Nonlinear response of $\beta$-Ta resonators}
\begin{figure}[h]
     \centering
     \includegraphics{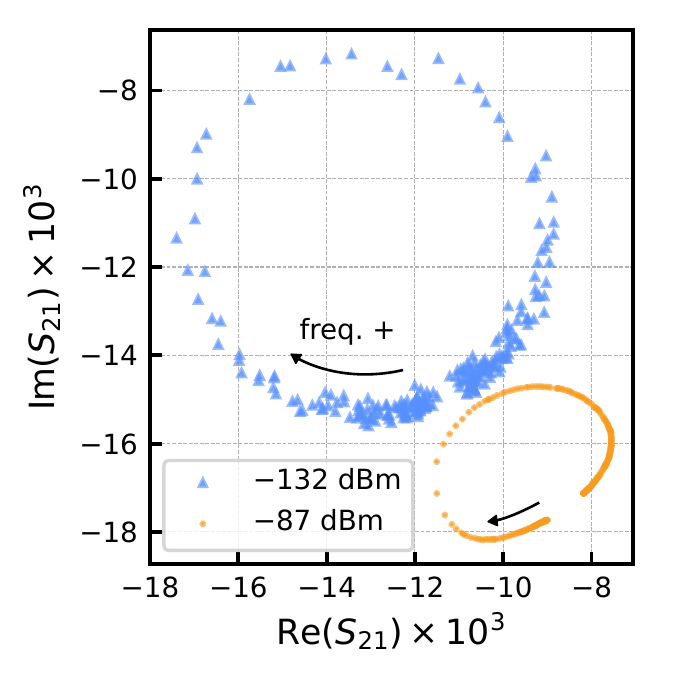}
     \caption{Trajectories of $S_{21}$ spectra in a complex plane. The data were taken for $\beta$-Ta resonator R4 with microwave probing powers of $-132\,\U{dBm}$~($\n\approx 5$) and $-87\,\U{dBm}$. The trajectory is circular at low probing power, while it becomes small and oval at high probing power owing to the nonlinearity of the superconducting film. The black arrows indicate the direction in which the frequency increases.\label{fig:beta-Ta-NL}}
\end{figure}

Figure~\ref{fig:beta-Ta-NL} shows the trajectories of the measured $S_{21}$ spectra of the $\beta$-Ta resonator in a complex plane.
At low probing power~($-132\,\U{dBm}$, corresponding to $\n\approx 5$), the trajectory has a circular shape, which is fitted well with Eq.~(\ref{eq:S21}). On the other hand, at high probing power~($-87\,\U{dBm}$), the trajectory becomes small and oval, and is no longer fitted well with Eq.~(\ref{eq:S21}). This behavior is attributed to the intrinsic nonlinearity of superconductors~\cite{zmuidzinas_superconducting_2012}. This is the reason why $\Qint$ of the $\beta$-Ta resonators is not plotted at large photon numbers in the other figures.

%
%
%

\section{Physical analyses}
Here, we summarize the results of supplementary physical analyses.
\subsection{XPS}
Figure~\ref{fig:XPS} shows the XPS results of the $\alpha$-Ta sample before and after the DHF treatment.
The measurement was performed with an XPS apparatus~(KRATOS Nova, Kratos Analytical) with a monochromatic Al K$\alpha$ source and an electron take-off angle normal to the sample surface.
The spectra of Si 2p were measured at the gap around the bonding pad for the CPW feed line in Fig.~\ref{fig:setup}(a), and those of Ta 4f were measured at the ground plane between R2 and R4.

In the DHF treatment, the sample was dipped in 50:1 diluted HF for 5 min and then rinsed with running deionized water for $>$5 min.
The sample was transferred into the load-lock chamber of the XPS apparatus within 40 min after the acid treatment.
The Si 2p spectra clearly show that Si surface oxides are removed by the DHF treatment.
On the other hand, the Ta 4f spectra suggest that Ta surface oxides remain and are more resistant to 50:1 DHF than Si surface oxides.

%
%
\begin{figure}[htb]
     \centering
     \includegraphics{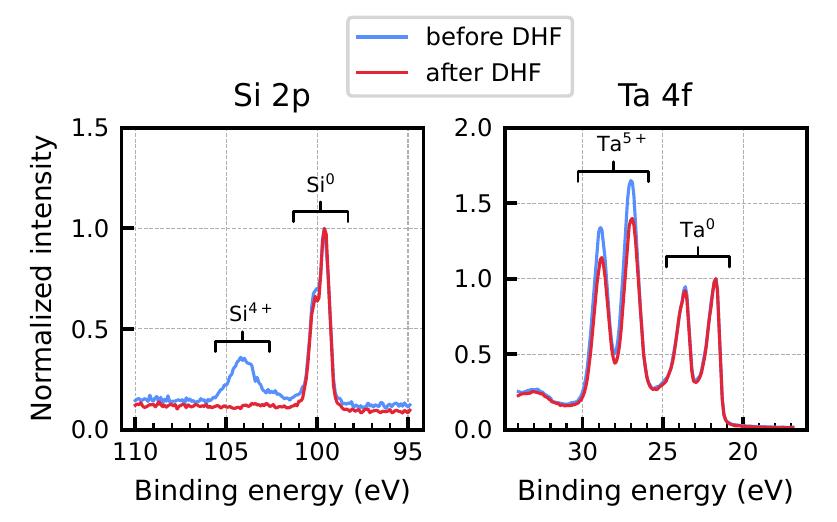}
     \caption{XPS spectra of the $\alpha$-Ta film before and after the DHF treatment. The left and right panels show the spectra for Si 2p and Ta 4f, respectively. The spectra are normalized to the intensity of the peaks at the lowest binding energy.\label{fig:XPS}}
\end{figure}

\subsection{HAADF-STEM and EELS}
\begin{figure}[htb]
     \centering
     \includegraphics{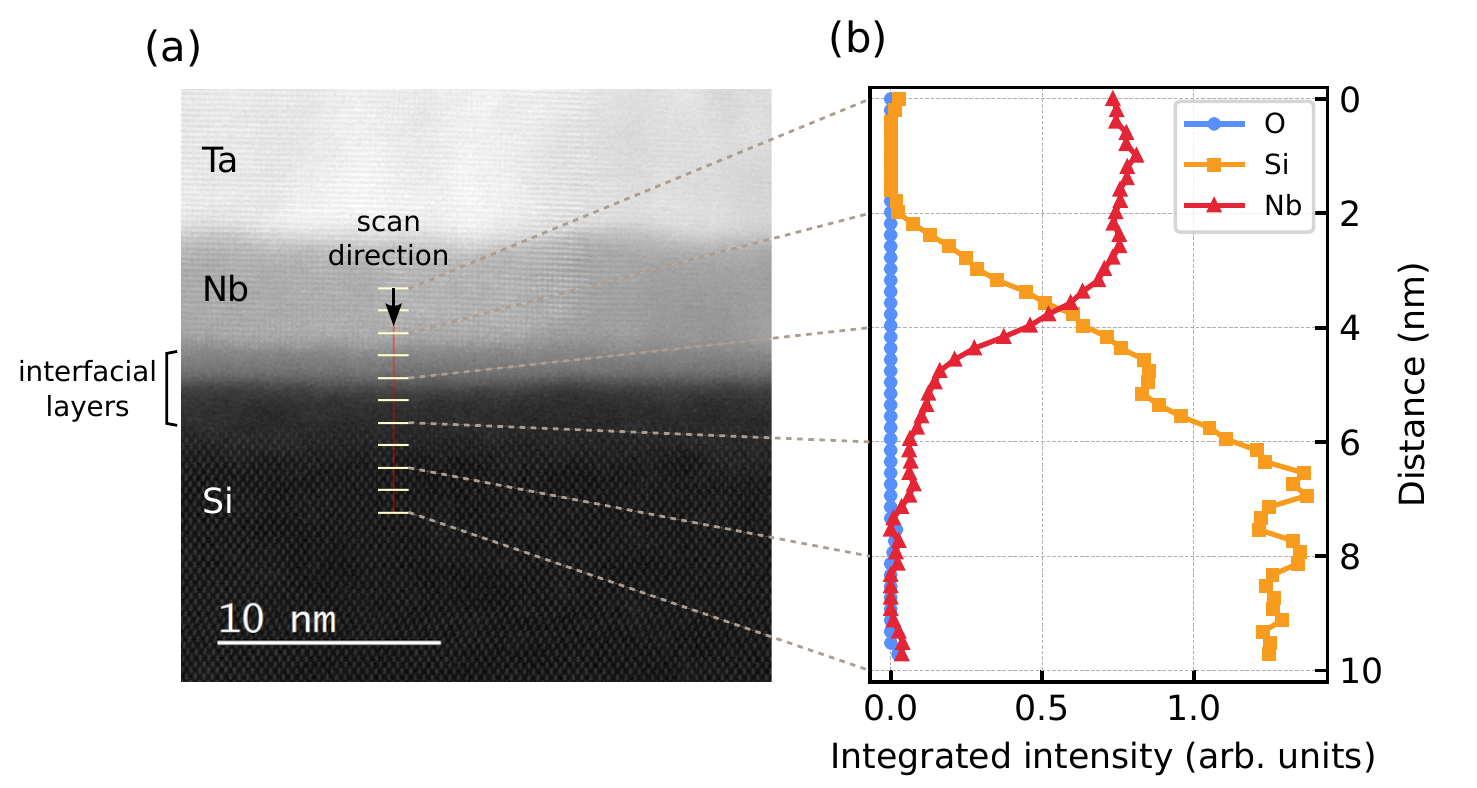}
     \caption{(a)~HAADF-STEM image of the interface between the Nb buffer layer and the Si substrate. Two amorphous interfacial layers can be seen. (b)~Integrated intensities of O-K, Si-K, and Nb-L$_3$ peaks in the EELS spectra along the scan line shown in (a).\label{fig:EELS}}
\end{figure}

Figure~\ref{fig:EELS}(a) shows a high-angle annular dark-field scanning TEM~(HAADF-STEM) image of the interface between the $\alpha$-Ta film and the Si substrate. We can see two amorphous interfacial layers.
These amorphous interfacial layers are considered to be the major causes of the TLS loss of the present resonators.

To investigate the interfacial layers, we performed electron energy-loss spectroscopy~(EELS) along the scan line shown in Fig.~\ref{fig:EELS}(a). Figure~\ref{fig:EELS}(b) shows the integrated intensities of the EELS peaks corresponding to O-K, Si-K, and Nb-L$_3$ along the scan line.
The results reveal that the upper interfacial layer mainly contains Nb, and the lower one mainly Si.
In addition, the oxygen signal is not significant. Therefore, we conclude that the lower interfacial layer is composed mainly of amorphous Si and the upper one is composed of amorphous Nb grown on the amorphous underlayer.

%
%

\section{Loss model of CPW resonators}
The total loss $\delta = 1/\Qint$ of the resonator is expressed as the sum of the losses due to the different causes as follows:
\begin{equation}
     \delta = \frac{1}{\Qint} = \delta\sub{TLS} + \delta\sub{QP} + \delta\sub{other},\label{eq:delta_tot}
\end{equation}
where $\delta\sub{TLS}$ and $\delta\sub{QP}$ denote the losses due to TLS and quasiparticles, respectively, and $\delta\sub{other}$ represents constant losses due to other causes such as radiation, sample package, and vortices.
\subsection{Loss originating from TLS}
The loss due to a TLS ensemble is represented by~\cite{gao_physics_2008}
\begin{equation}
     \delta_\mathrm{TLS} = \frac{F\delta_\mathrm{TLS,0} \tanh\left(\frac{\hbar \omega}{2 k_\mathrm{B}T}\right)}{\sqrt{1 + \frac{\Omega^2}{\varGamma_1 \varGamma_2}}},\label{eq:delta_TLS_org}
\end{equation}
where $F$ is the participation ratio of the electric energy in the volume of interest; $\delta_\mathrm{TLS,0}$ is the intrinsic TLS loss in the zero photon and temperature limit; $\hbar$ is the reduced Planck constant; $\omega$ is the angular frequency; $k_\mathrm{B}$ is the Boltzmann constant; $T$ is the temperature; $\Omega$ is the parameter related to the electric field strength in the resonator and $\n$; and $\varGamma_1$ and $\varGamma_2$ are the average energy-relaxation and dephasing rates of the TLS ensemble, respectively.

According to Ref.~\cite{crowley_disentangling_2023-1}, we have
\begin{align}
     \varGamma_1 & \propto 1/\tanh\left(\frac{\hbar \omega}{2 k_\mathrm{B}T}\right), \label{eq:Gamma1} \\
     \varGamma_2 & \propto T^{\beta_1},
\end{align}
where $\beta_1$ is an exponent accounting for the temperature dependence of the dephasing rate owing to TLS--TLS interactions~\cite{burnett_evidence_2014,crowley_disentangling_2023-1}.
In addition to the model in Ref.~\cite{crowley_disentangling_2023-1}, we empirically introduce the lower bound of $\varGamma_2$ as
\begin{equation}
     \varGamma_2 = \varGamma\sub{2,min} + \gamma T^{\beta_1}.\label{eq:Gamma2_with_min}
\end{equation}
This modification is reasonable from the experimental standpoint because there is an incomplete thermalization of the TLS ensemble in the actual system.

Substituting Eqs.~(\ref{eq:Gamma1}) and (\ref{eq:Gamma2_with_min}) into Eq.~(\ref{eq:delta_TLS_org}) and simplifying the terms, we obtain
\begin{equation}
     \delta_\mathrm{TLS} = \frac{F\delta_\mathrm{TLS,0} \tanh\left(\frac{\hbar \omega}{2 k_\mathrm{B}T}\right)}{\sqrt{1 + \left(A+{C}{T^{\beta_1}}\right)^{-1} \tanh\left(\frac{\hbar \omega}{2 k_\mathrm{B}T}\right)}}, \label{eq:delta_TLS}
\end{equation}
where $A$ is a term originating from $\varGamma\sub{2,min}$ and $C$ is a coefficient in units of $\U{K}^{-\beta_1}$ and proportional to $\gamma$.
Note that $A$ and $C$ depend on $\n$.

\subsection{Loss originating from thermal quasiparticles}
Here, we consider the loss originating from thermal quasiparticles in superconductors. On the basis of the \MB{} theory~\cite{mattis_theory_1958}, the complex conductivity $\sigma = \sigma_1 + \jj\sigma_2$ of superconductors is approximated, when $\hbaro,\,\kBT \ll \Delta_0$, as~\cite{gao_equivalence_2008, zmuidzinas_superconducting_2012}
\begin{align}
     \frac{\sigma_1}{\sigma_\mathrm{N}} & = \frac{4\Delta_0}{\hbaro}\exp\left(-\frac{\Delta_0}{\kBT}\right) \sinh \left( \frac{\hbaro}{2\kBT} \right) K_0 \left( \frac{\hbaro}{2\kBT} \right), \label{eq:sigma1} \\[10pt]
     \frac{\sigma_2}{\sigma_\mathrm{N}} & = \frac{\pi\Delta_0}{\hbaro}, \label{eq:sigma2}
\end{align}
where $\sigma_\mathrm{N}$ is the normal-state conductivity; $\Delta_0$ is the superconducting gap energy at zero temperature; and $K_0(x)$ is the zeroth-order modified Bessel function of the second kind.
From the Bardeen--Cooper--Schrieffer~(BCS) theory~\cite{bardeen_theory_1957}, $\Delta_0 \approx 1.76 \kBT\sub{c}$.

The internal loss of a CPW resonator due to thermal quasiparticles, $\delta\sub{QP}=1/Q\sub{QP}$, is given by~\cite{zmuidzinas_superconducting_2012,de_visser_quasiparticle_2014}
\begin{align}
     \delta\sub{QP} & = \frac{\alpha\sub{k}\beta}{2} \cdot \frac{\sigma_1}{\sigma_2}\notag                                                                                                            \\
                    & = \frac{2 \alpha\sub{k}\beta}{\pi} \exp\left(-\frac{\Delta_0}{\kBT}\right)\sinh \left( \frac{\hbaro}{2\kBT} \right) K_0 \left( \frac{\hbaro}{2\kBT} \right),\label{eq:delta_QP}
\end{align}
where $\alpha\sub{k}$ is the kinetic-inductance fraction defined by the ratio of the kinetic inductance to the total inductance and $\beta := 1 + (2d/\lambda)/\sinh(2d/\lambda)$ is an order-unity factor accounting for the finite thickness of the superconducting film. Here, $d$ is the film thickness and $\lambda:=\sqrt{1/(\mu_0 \omega \sigma_2)}$ is the penetration depth, where $\mu_0$ is the permeability of vacuum. For the present $\alpha$-Ta film, we have $\lambda = 121\,\U{nm}$ from the measured resistivity and $T\sub{c}$.

\bibliographystyle{../aipnum4-1}
\bibliography{../Ta_CPWR_paper.bib}